\newcommand{\beq}{\begin{quote}}

\newcommand{\enq}{\end{quote}}
\newcommand{\be}{\begin{equation}}
\newcommand{\en}{\end{equation}}

\documentclass[prb,preprint]{revtex4-1} 
\usepackage{amsmath}  % needed for \tfrac, \bmatrix, etc.
\usepackage{amsfonts} % needed for bold Greek, Fraktur, and blackboard bold
\usepackage{graphicx} % needed for figures

\begin{document}

\title{Theory and experiments  on the ice-water front  propagation in droplets 
freezing on a subzero surface.  }

\author{Michael Nauenberg}
\email{michael@physics.ucsc.edu} % optional
\affiliation{Department of Physics, University of California , Santa Cruz, CA 95064}

\date{\today}

\begin{abstract}
An  approximate theory  is presented describing   the propagation of the ice-water front that develops  in 
droplets  of water that are deposited on a planar surface at a temperature below the melting point of ice.  A calculation based on this
theory  is  compared with our   experimental observations of the time evolution of this front.  The results of  calculations of
this  front by Schultz {\it et al}  \cite{schultz}, obtained by  integrating numerically the exact 
differential equations  for this problem,  were published graphically, but  only for the time-dependent velocity of
this front. Unfortunately, these theoretical results
cannot be compared directly with our experimental observations. Our experiments were performed
by  freezing  water droplets  directly on a block of dry ice, and in order to examine the  effects of the heat conductivity of a 
substrate during the freezing process, such  droplets were also deposited 
  on a glass plate and on a copper plate  placed on dry ice. The  temperature at the base  of 
  these droplets, and  the dependence of the freezing time on their size  was   also investigated experimentally, and 
  compared with our analytic approximation of the theory.  Such  experiment have not been published previously,  and
  reveal that the usual assumption that the temperature at the base of the droplets
  is a constant,  made in  all previous theoretical papers on this subject,  cannot be implemented in practice.
 
 \end{abstract}

\maketitle

\section{Introduction}
Recently, there has been some  renewed interest  in the properties of  water  droplets that are frozen on a planar surface
  at  subzero temperatures \cite{anderson, snow, oscar, michael,schultz, jacco,michael2}. A  theoretical analysis, and numerical solutions 
  of the heat diffusion differential  equations  for the  propagation
of the ice-water front in spherical droplets have  been carried out 
by  W.W. Schultz, M. G. Worster, and D.M. Anderson \cite{schultz}, under the assumption that the temperature
at  the base of the droplet in direct  contact with this surface is a constant,   but up to the present time a comparison of the theory with 
observations of this propagation have not been reported.  I present  here  the results of  such observations with droplets deposited 
on a block of dry ice, and on glass and copper plates placed on this block.  It turns out that the heat conductivity of
these plates  is very important in establishing  the temperature at the base of the droplet, and the rate of freezing.
To make the ice-water boundary  easily observable,
a drop of   food coloring was added to a glass of  tap 
water used in the experiment.  While it is   dark blue in the liquid phase, it becomes  green in
 the frozen phase, see  Fig.1.   The time-dependent   height  $h$ of this boundary, ending in a fully frozen droplet,  Fig.2,
 was recorded with  a video camera,   and the results are presented graphically  in Sec. II, Fig.6,   where 
$h$   is plotted as a function of the  time $t$.   A curve  fitting  the experimental data
on this plot, is based on an 
approximate theory of this freezing process discussed in Section I.

\maketitle 
\begin{figure}[ht]
\centering
\includegraphics[width=10cm]{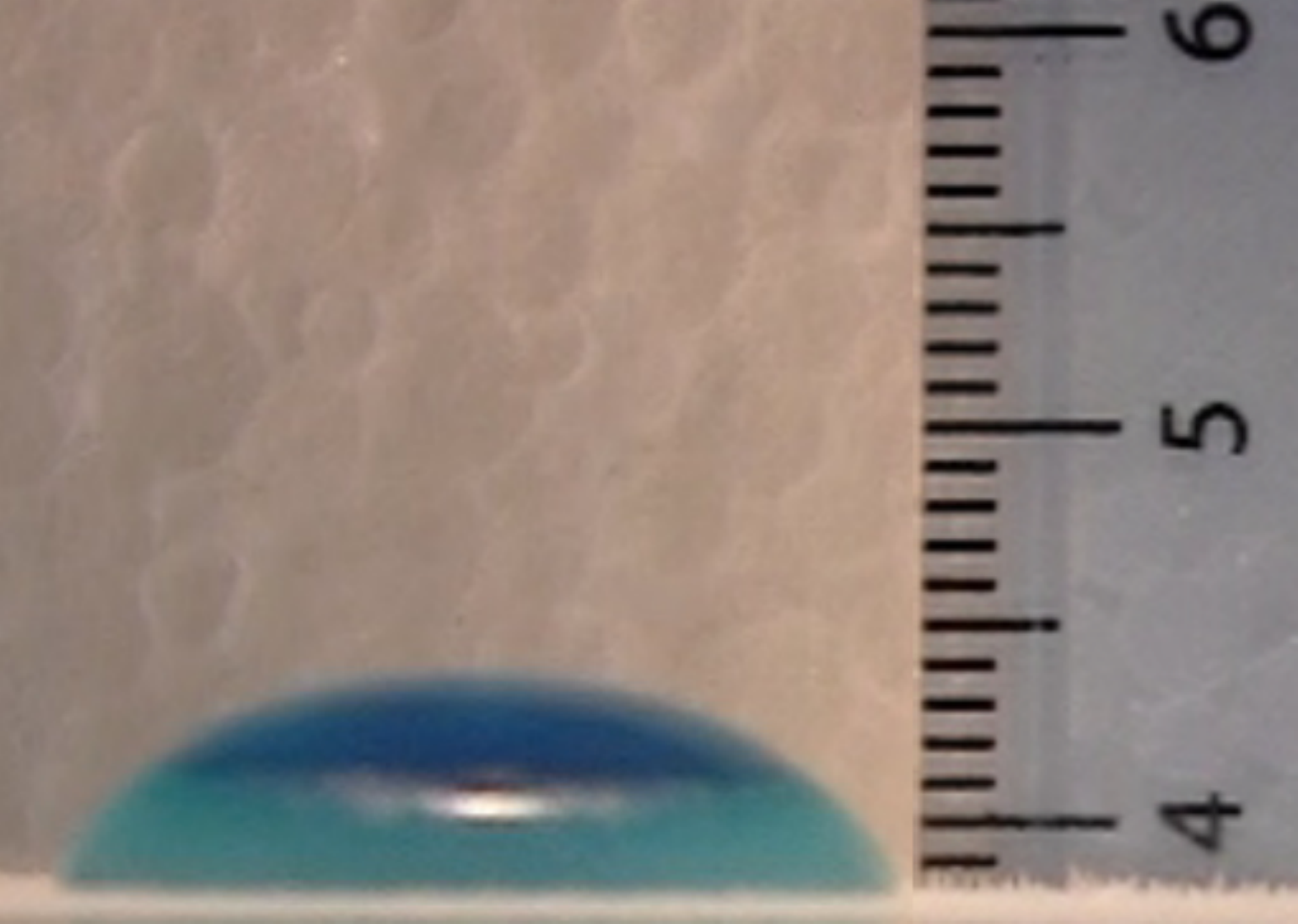}
\caption{Droplet freezing on  a glass plate placed on a  block of dry ice.
The ice-water front is at the boundary between the dark blue and the light
blue regions for the water and the  ice phase, respectively.
} 
\label{}
\end{figure}
\pagebreak
\maketitle 
\begin{figure}[h]
\centering
\includegraphics[width=10cm]{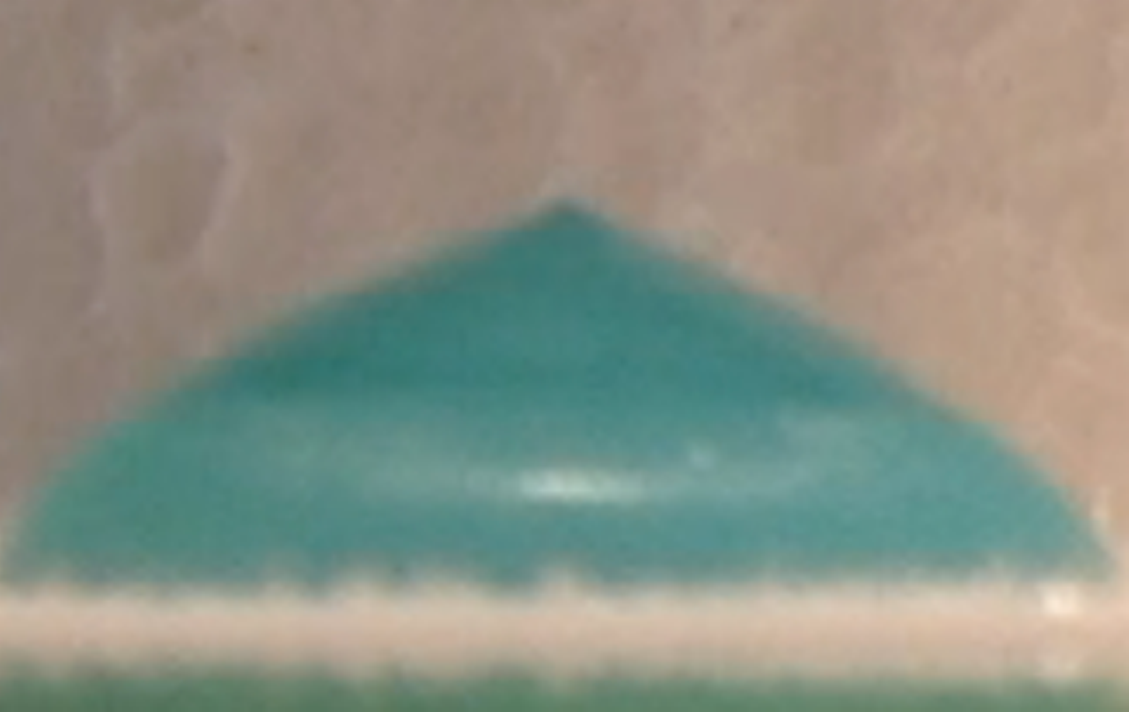}
\caption{Droplet shown in Fig. 1 after it is completely  frozen.
} 
\label{}
\end{figure}

When the liquid phase  is  
at the melting temperature of ice,  the theory for  the propagation of a planar ice-water front  
 was  given a long time ago by Joseph Stefan\cite{stefan}.
Neglecting the heat capacity of ice, the height of the front as a function of time  is
 $h(t)=\sqrt{c(t) t}$, where $c(t)=2k_i \overline T(t)/L $,
 $k_i$ is the  heat conductivity of ice, $L$ is the latent heat (per unit volume), and   $\overline T(t)$ is 
 the average  magnitude,  during an interval of time $t$,  of the temperature (in degree Celsius) below melting at the base of the droplet, 
 $\overline T(t)=(1/t)\int_0^t dt T(t)$.   
For example, the data shown in Fig.6 can be approximately fitted,  for $t \leq 100$ seconds, with a constant  value 
for  $c\approx .003\; cm^2/sec$, corresponding
to  $T \approx -21^o C$. The magnitude of this temperature  appears to be  surprisingly small because the temperature of dry ice   
 is $-78.5^o$ C,  but  due to  the transfer of  latent heat during   the freezing process, it is expected that  
 the droplet heats the dry  ice surface underneath it.  The effect of this heating  can be observed, for example,  in the appearance  
 of a small cavity in the dry ice  after  the fully frozen droplet  is removed.
  In Section II, direct measurements
of the temperature at the base of several droplets are presented that  verify this heating process,  and  its  dependence on time.  
For $t \geq 100$ 
seconds, the height of the ice-water  front  increases more rapidly as a function of time than expected
from Stefan's relation. This increase in the velocity can be understood, because  the
 approximation that the ice-water front is nearly flat, required by the application of Stefan's relation, 
 ceases to be valid, and  this front becomes rapidly concave.
This change in shape has been observed experimentally \cite{michael, snow2,jacco},  and it is also found
 in the numerical calculations of Schultz et al \cite{schultz}.  Under the assumption that the shape of this front can be
approximated by a spherical cap of constant curvature that is  normal to the surface of the droplet, the
 top of the  the droplet  takes the shape of a cone\cite{schultz, jacco}.  A novel feature in my theoretical  approach,  discussed in  Section I,
 is to treat analytically  the propagation of such a spherical ice-water front by an  extension of Stefan's relation for a planar front.
 In Section II  a calculation based on this approach   is compared with  an experimental observation of  the front propagation in a droplet
 deposited on dry ice. In section III,   I present some time dependent measurements of the
 temperature at the base of droplets of various sizes  on dry ice, and also on a glass plate and on copper plate. 
 Some of my conclusions are summarized in  Section IV.
 
 \begin{figure}[h]
\centering
\includegraphics[width=15 cm]{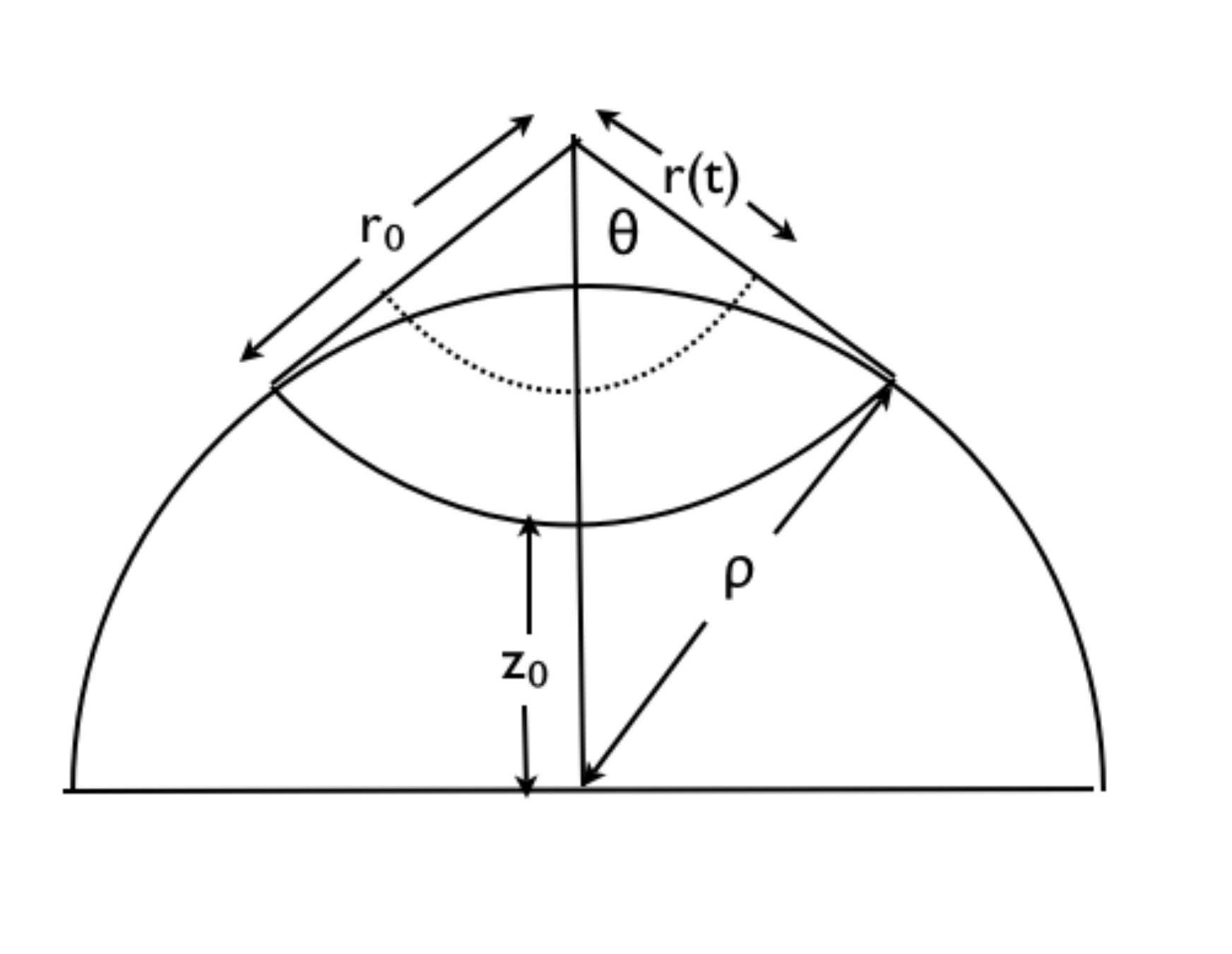}
\caption{Propagation of a hemispherical ice-water front, indicated  by the circular dashed line  with radius $r(t)$,  centered at the apex
of  a cone with opening angle $\theta$. In the final stages of freezing, this cone becomes   the shape near
the top of the frozen  water drop with initial radius $\rho$.}. 
\label{}
\end{figure}

%\section*{ I. Analytic approximation for the propagation of the  ice-water front. }
\section*{ I. Analytic approximation for the propagation of the  ice-water front. }
In this section,  I present  an  analytic calculation of the ice-water front propagation in  a hemispherical  droplet of water
placed on a planar surface at a temperature below the melting point of ice. 
This calculation is based  on the  approximation that  the propagation of this front 
can be  split into two stages: a)  the front 
is planar until it reaches a critical height when the shape of the droplet  evolves into a cone due to the expansion of the
volume of water when it freezes into ice, and
 b) afterwards,  the front takes  the shape of  a spherical cap  centered at the  apex of this cone \cite{schultz, jacco}, 
  illustrated  in Fig. 3. 
 For simplicity we assume that the temperature $T_i$  at  the base of the droplet  is constant, but the theory can be
 readily generalized to account for the experimentally observed time dependence of $T_i$  presented in section III.

Provided that  the liquid  phase of the droplet   remains at the melting temperature $T_m$,  the heat flux generated by the
latent heat that emerges  during freezing is conducted primarily by the ice phase of the droplet. 
 Then the  propagation velocity $v$  of a planar ice-water front moving along the positive $z$ axis  is  determined by the 
 energy conservation relation\cite{stefan}
\be
\label{stefan}
vL=k_i\frac{\partial T}{\partial z},
\en
where $T$ is the temperature at the front, $L$ is the latent heat per unit  volume, and $k_i$ is the heat conductivity
of ice.  Neglecting the heat capacity of ice, $\partial ^2 T/\partial z^2=0$, and   $T$ depends linearly  on $z$,
with $T=T_i$ at $z=0$,  where the drop is in contact with a subzero heat reservoir, and $T=T_m$ at the propagating
ice-water front located at $z=z(t)$. Hence
\be
\label{temp1}
T=T_i+(T_m-T_i)\frac{z}{z(t)},
\en
and  Eq.\ref{stefan} becomes  an equation of motion for the propagation of the ice-water front, 
\be
\frac{dz(t)}{dt}=\frac{k_i(T_m-T_i)}{Lz(t)}.
\en
Setting $z'(t)=z(t)/\rho$, where $\rho$ is a length scale that we take to be the radius of the drop, and $t'=t/t_1$, where $t_1=L \rho^2/2k_i (T_m-T_i)$ is a time scale,
one  obtains Stefan's relation in the dimensionless form
\be 
t'=z'^2.
\en

The  main approximation  now is to assume that this relation holds until $z=z_o$, where  $z_o$ 
is the location  of the ice-water front  when it  attains  the shape of a  spherical cap. This cap  is centered at the 
cusp of a cone with  opening angle $\theta$,
and side length $r_o=\rho cos\theta/sin\theta$, and $z_o  =\rho sin\theta-r_o(1-cos\theta)=\rho sin\theta/(1+cos\theta)$ 
 (see Fig. 3). The angle $\theta$ is obtained  by the requirement that 
upon freezing, the remaining volume  of water  expands  into this cone and  into the
spherical cap \cite{schultz,jacco}. The  ratio $\nu=.917 $ of ice to water density 
determines $\theta\approx 65^o$,  and   corresponds to $z_0\approx .64$.
Assuming that for   $z\geq z_o$ the ice-water front has the shape of a hemisphere centered at the cusp of the cone,  
\be
v_r L= -k_i \frac{\partial T}{\partial r},
\en
where  the origin of the radial distance $r$ is at the apex of the cone of length $r_o$,  the radial velocity of the ice-water front is $v_r=dr(t)/dt$,
and the temperature $T$ satisfies the spherically symmetric Laplace equation
\be
\frac{\partial^2T}{\partial r^2}+\frac{2}{r}\frac{\partial T}{\partial r}=0.
\en
Hence $ T=\alpha/r +\beta$, where $\alpha$ and $\beta$ are time dependent variables.  I determine these variables  by the condition that
at $r=r(t)$, $T=T_m$, while at $r=r_o$, $T=T_o$, where according to Eq.\ref{temp1}, when $z(t)=z_o+r_0-r(t)$, 
\be
T_o= T_i+\frac{(Tm-Ti)z_o}{z_o+r_o-r(t)}.
\en
Hence  $\alpha=(T_m-T_o) r(t)r_o/(r_0-r(t))$, and  the gradient of $T$ at the spherically  shaped  ice-water front at $r=r(t)$ is
\be
\frac{\partial T}{\partial r}=-\frac{(T_m-T_i)r_o}{(z_o+r_o-r(t))r(t)}.
 \en
 As a bonus,  one finds that when $r(t)=r_o$  the linear and the spherical gradient of the temperature $T$ are the same. Therefore, at
 this cross-over,  the velocity of the front is continuous.
 Setting $x=r(t)/r_o$, and $t''=t/t_2$, where  $t_2=r_o^2 L/2k_i(T_m-T_i)$ is the relevant time scale for the
 propagation of the spherical ice-water front, 
 \be
 \frac{d x}{dt''}=-\frac{1}{2(\eta+1-x)x},
 \en
where $\eta=z_o/r_o=1/cos\theta - 1$. Integrating this equation of motion, 

\be
t''(x)=t''(1)+(\eta+1/3)-x^2(\eta +1 -2x/3),
\en
where $t''(1)=(t_1/t_2)sin^2\theta/(1+cos\theta)^2$,   and $t_1/t_2=\rho^2/r_0^2= sin^2\theta/cos^2\theta$.
When $x=0$, the ice-water  front has reached the apex of the cone, and the freezing  is completed, with  the elapsed 
scaled time  
\be
t'_f=\frac{1}{1+cos\theta}+\frac{cos^2\theta}{3sin^2\theta}.
\en
For $\theta=65^0$, $t'_f=.775$, Fig.4 shows  the time dependence of the height $z(t)$ of the ice-water front  
 in scaled variables, and  Fig.5 shows
 the velocity $v(t)$ of this front.   The transition from a planar to a spherical ice-water front 
occurs at about half the total freezing time  $t_d=z_0^2\approx .4$. The position and velocity of the front
are continuous at this point,  but  in our approximation  that at this time  the ice-water front changes discontinuously
from a planar to  a spherical front,   this effect can be seen as  a small discontinuity in the acceleration
of this front .  

When the temperature at the base of the droplet is not a constant, but  depends on the freezing  time $t$, which in practice
is  the case,  as shown in
Section III , the scaling factors $t_1$ and $t_2$ become function of the time.  Then $t_1(t)=L \rho^2/2k_i (T_m-\overline{T_i(t)})$,
and $t_2(t)=L r_o^2/2k_i (T_m-\overline{T_i(t)})$, where $\overline {T_i(t)}$=    $(1/t)\int_0^{t} dt' T_i(t')$ is the
mean temperature at $t$ for $0 \leq t \leq t_o$,
and $\overline {T_i(t)}$=    $(1/t)\int_{t_o}^{t} dt' T_i(t')$ is the mean temperature for $t_o \leq t$.

\begin{figure}[b]
\centering
\includegraphics[width=12cm]{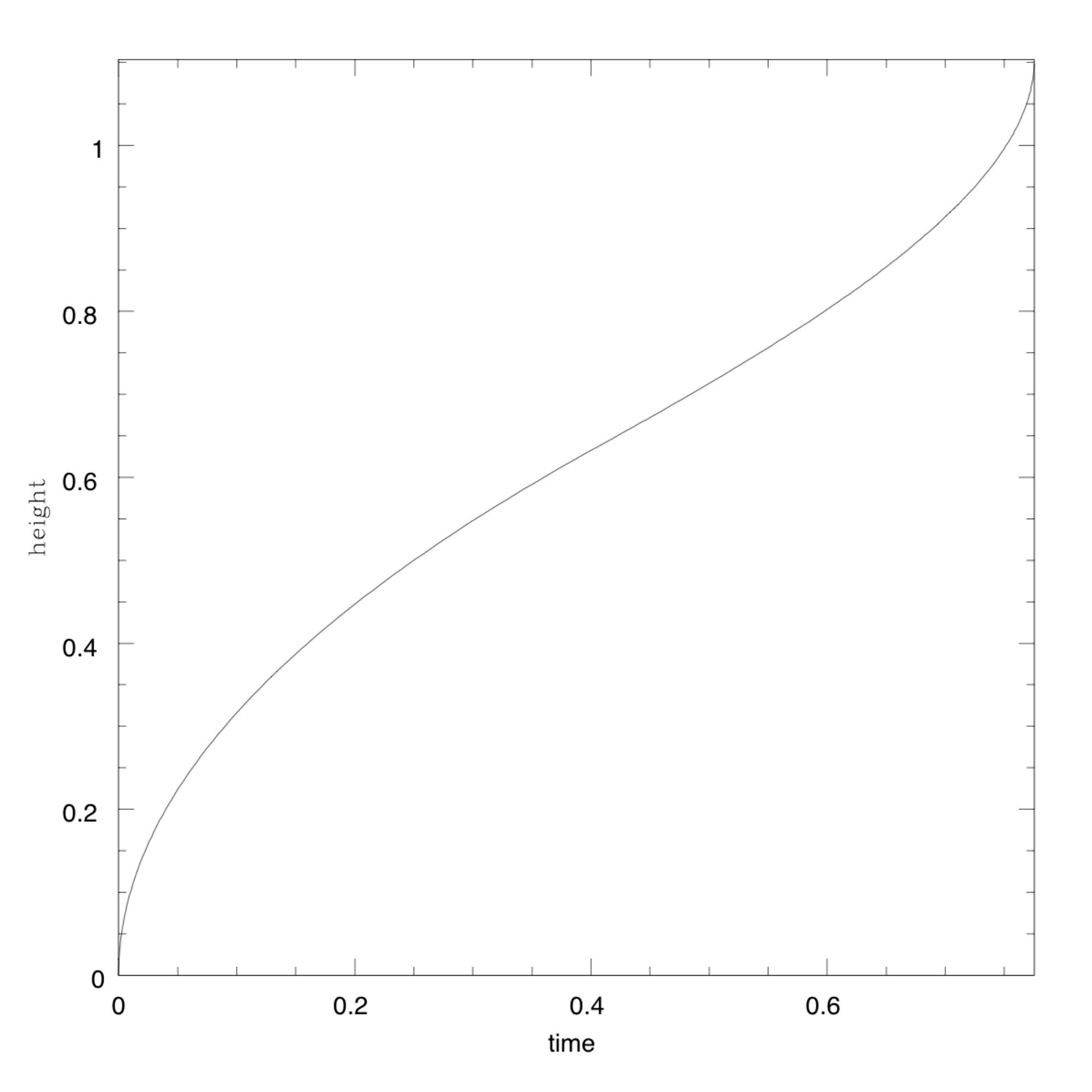}
\caption{Height   of the ice-water front vs. time in scaled variables, for
a drop of water freezing on a surface held at a constant temperature below the melting temperature of ice.}
\label{}
\end{figure}

\begin{figure}[h]
\centering
\includegraphics[width=12cm]{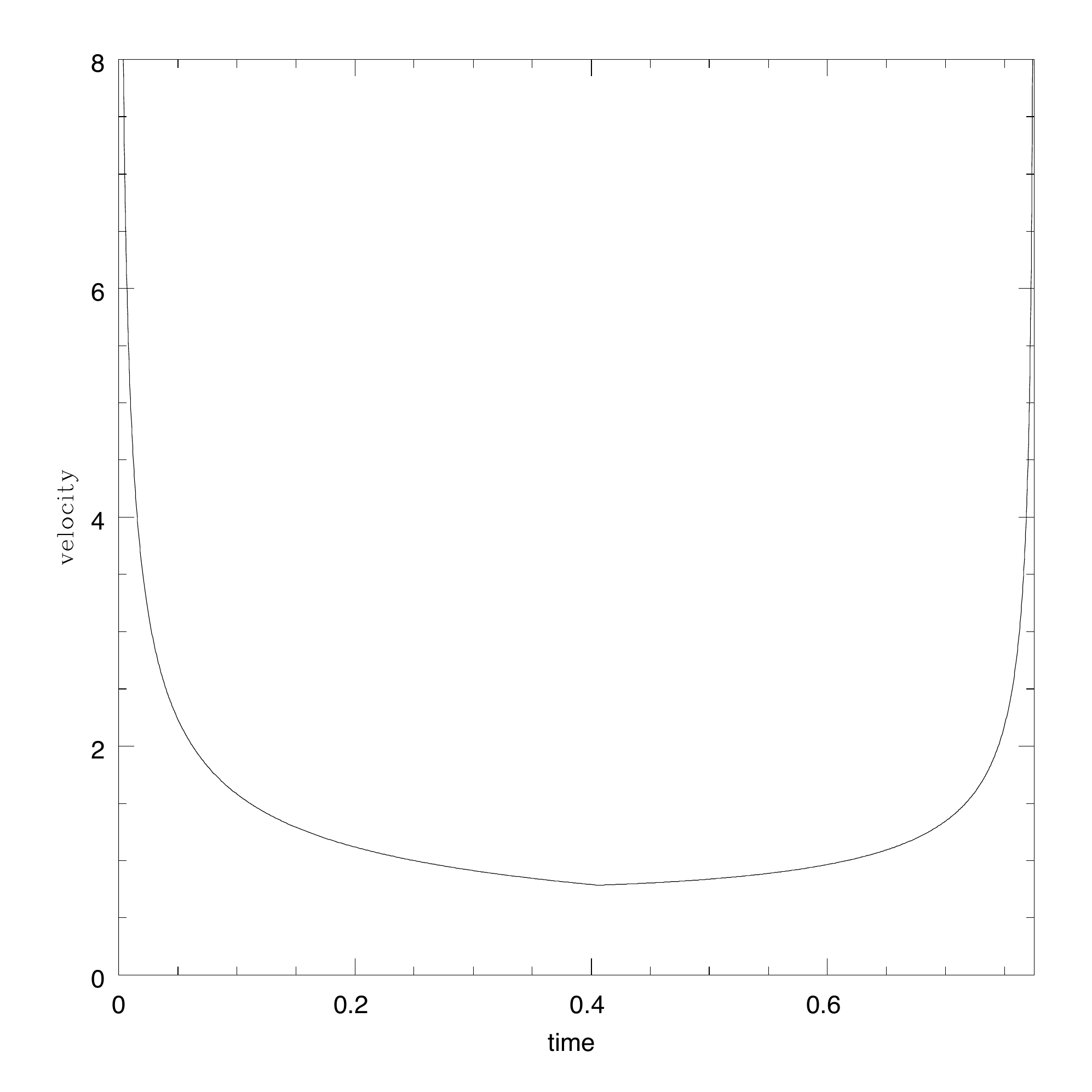}
\caption{Velocity  of the ice-water front vs. time in scaled variables, for
a drop of water freezing on a surface held at a constant temperature below the melting temperature of ice.}
\label{}
\end{figure}

\clearpage

%\section*{ II. Experimental results and comparison with the theory}

\section*{ II. Experimental results and comparison with the theory}

The  propagation of the ice-water front was observed  in  droplets  of water  deposited on a glass
surface placed on a block of dry ice.  An example is  shown in  Figs.1 and 2, and
the height  of the ice-water front, recorded  with a video camera (Cannon Vixia HFS200),  is plotted as a function of the 
time in Fig.6.  The  curve shown on this plot  is a fit to this data based on the theory
discussed in the previous section. The main source of errors occur in the determination of the height of the ice-water front at a given time. As can be be seen in Fig 1, this front , revealed by the change of colors between the ice and water phases in the droplet, is not completely sharp. But Fig.1, which also shows in the margin the scale in millimeters, indicates that the error is only a fraction of a millimeter.   

The bulk temperature of the dry ice is  $  -78.5$ degrees Celsius,  but in order 
to conduct the latent heat produced during the freezing process,  the temperature
at the contact between the base of the drop and the glass plate  must be higher.  Direct evidence for this heating was  observed by depositing the
drop directly on the surface of the dry ice. After the freezing process was completed,  removal of the frozen drop
left a small hole due to the heating  of the dry ice under the base of the drop.   In the next section, it  will be shown that this temperature is not a constant as has been  assumed in all the  theoretical work up to the present time,  but  for large size droplets 
its value  can be approximated by an average temperature during the freezing process.  
For example, in the  fit of the  analytic approximation discussed in Section I to the experimental data in Fig.6, 
the only  unknown parameter is  the  mean value of the temperature, given by  $T_i=- .39\rho^2 L/ k_i t_f $.   For  the
effective radius of the droplet,  $\rho \approx.7$cm., and the observed time to complete the freezing  process,  $t_f=120 $ seconds,  
 $T_i =  -22^o$C.  Unfortunately, this data  cannot  be compared also with the numerical calculations of 
 Schultz et al. \cite{schultz}, because 
 their  time dependent  results   were given only  for the velocity of  the ice-water front. \\

 \begin{figure}[h]
\centering
\includegraphics[width=15cm]{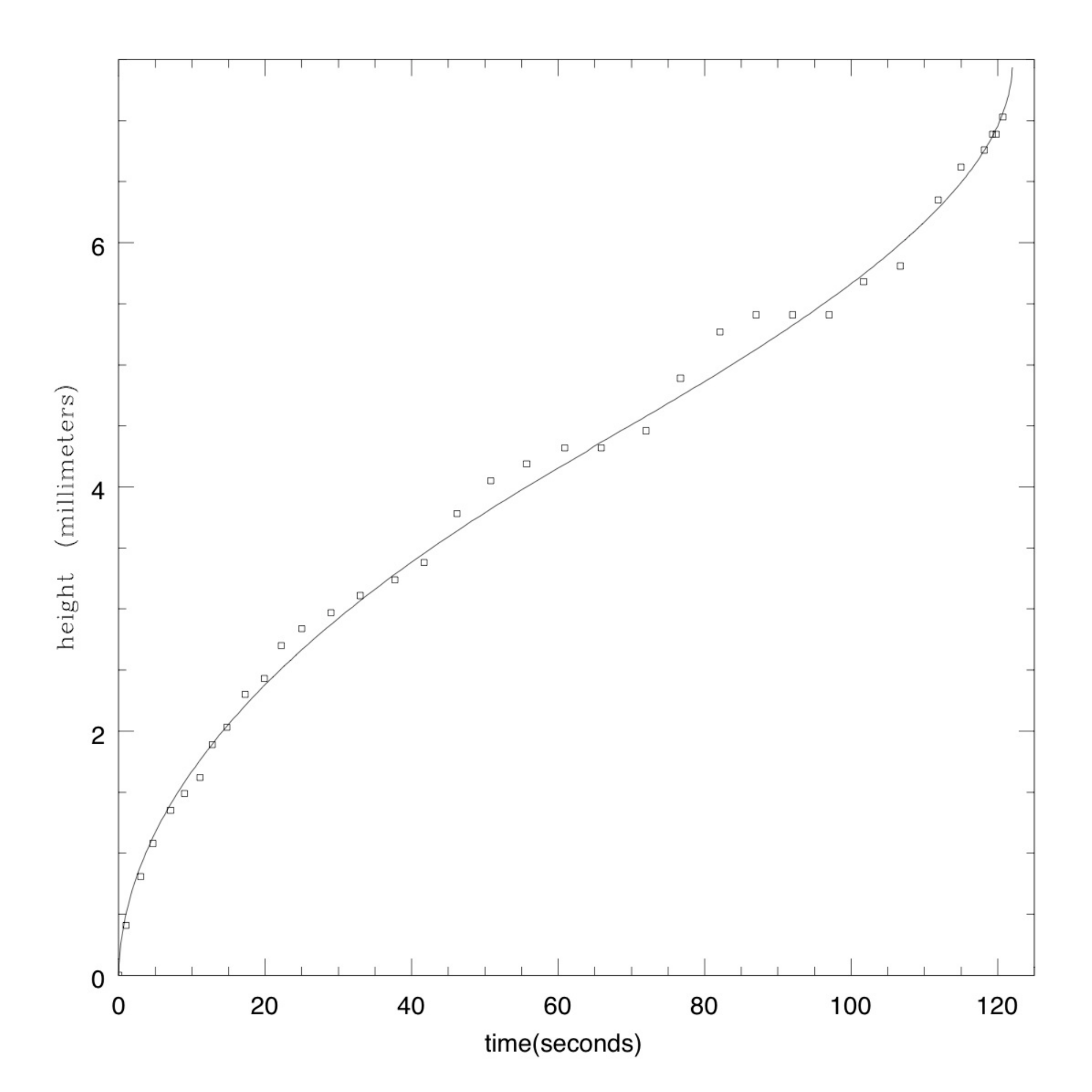}
\caption{The squares represented  the observed height   of the ice-water front vs. time  for
the water drop shown in Fig.1. The theoretical curve,  shown for scaled variables in Fig.4, 
fits  the experimental data  for  $T_i\approx -25^o$ C. }
\label{}
\end{figure}

\clearpage
% \section*{III. Temperature measurements}
  \section*{III. Temperature measurements} 
 
   During the freezing process, the temperature was measured  at the base of several droplets of different sizes deposited  on a block of dry ice,  Fig.7,  and  on  a  copper  plate .1 cm. thick,  placed on this block, Fig.8. These measurements were made with a thermistor $.2$ cm. in diameter. The main source of errors are due to this finite size of the thermistor, and to the heat generated by the current flowing through it.
   After the freezing process was completed, I removed the droplet from the substrate, and observed that the thermistor remained located at its base. In an extremely short time, that could not be determined in this experiment,
after these  droplets were deposited on these subzero surfaces, the temperature increased very rapidly to higher values that dependent on the size of the droplet, and afterwards it  decreased at  a rate  that was  slower  with increasing size of the droplet.  
 For the smallest droplet (.1 cm in radius) deposited on dry ice,  the decrease 
in temperature was very  sharp,  but  for the larger  drops the assumption of a nearly constant temperature during the  freezing process 
made in the numerical calculations of Schultz {\it et al.} \cite{schultz}, and  in my analytical calculations in Section I,  
should be applicable.  The inflection that appears  at longer times  in the temperature vs. time curve, Fig.7,  occurred  when the freezing of the drop was nearly completed,  and the release of latent energy had ended. But as expected, 
 the temperature of the frozen drop continues to  drop until it came into thermal equilibrium with the dry ice. 
 A similar behavior was observed  with droplets on the copper plate, Fig.8, but in this case the rate of freezing  was approximately
 10 times faster.  In this case, there appeared  also some unexpected jumps in the temperature as function of time at the end
 of the freezing process  that remain unexplained .  \\

 \begin{figure}[h]
\centering
\includegraphics[width=15cm]{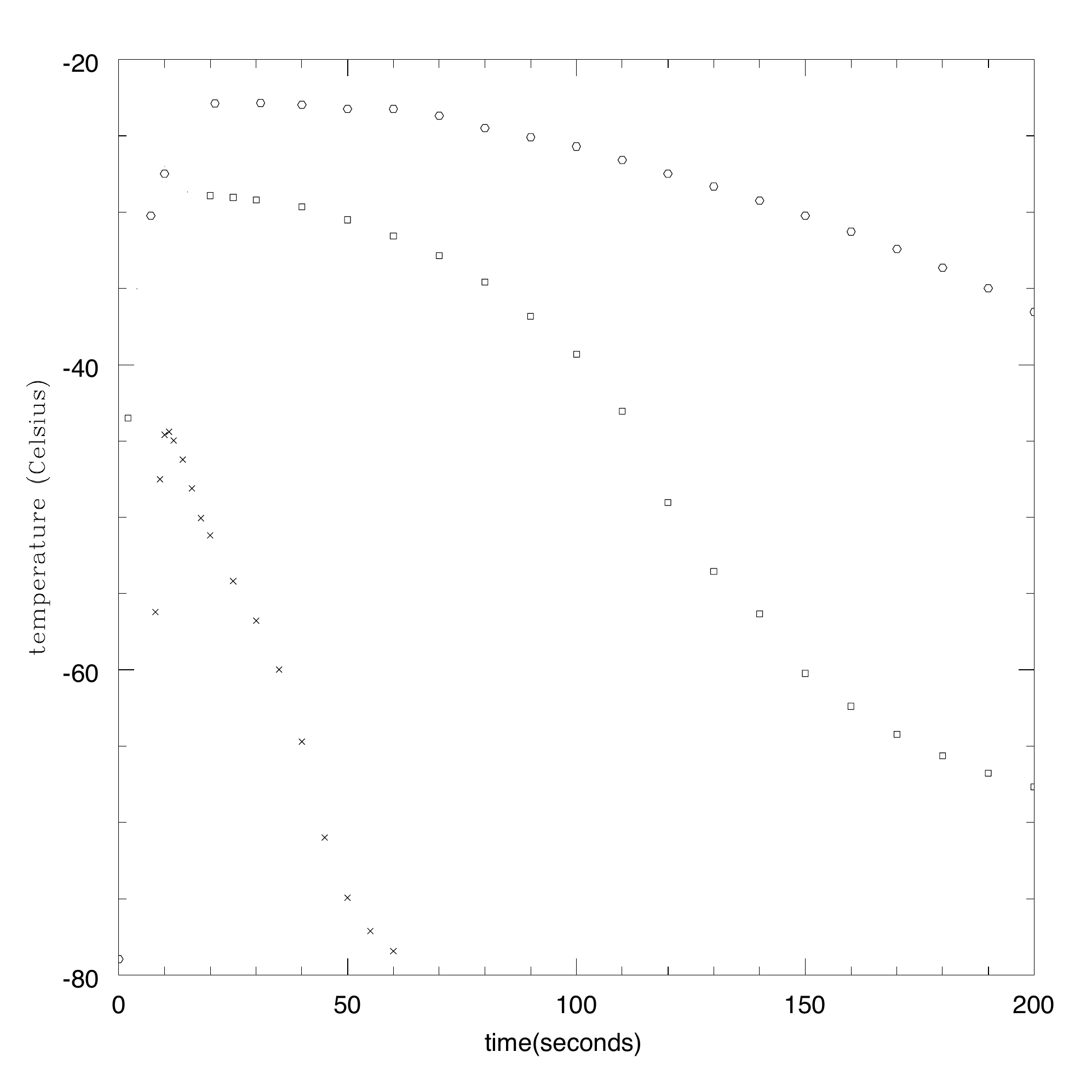}
\caption{The temperature  at the base of droplets of water of different sizes deposited  on a block of dry ice,
as a function of time: $\rho=.23$ (crosses), $\rho=.5$ (squares) and $\rho=.63$ (hexagons)}
\label{}
\end{figure}

 \begin{figure}[h]
\centering
\includegraphics[width=15cm]{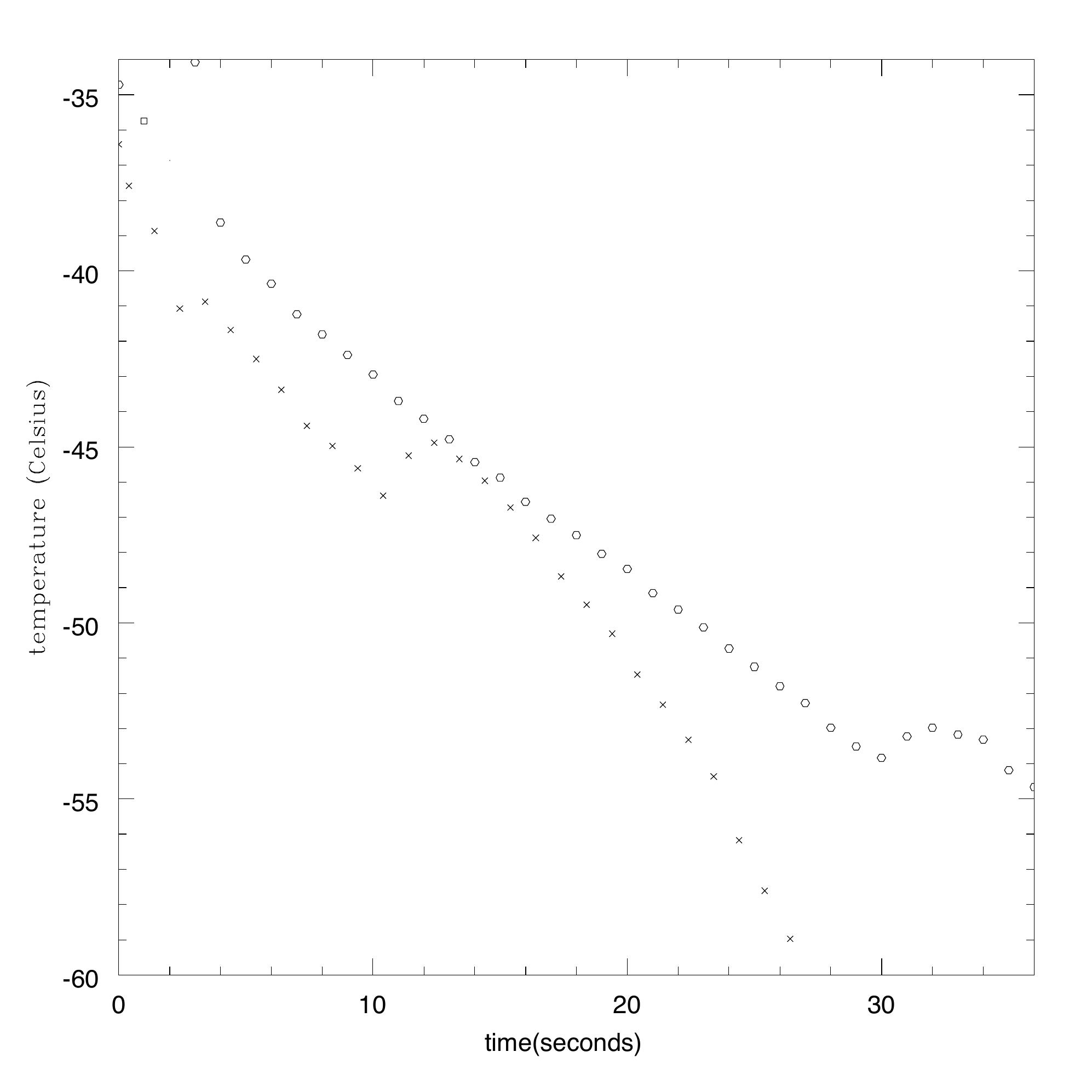}
\caption{The temperature  at the base of two droplets of water of different sizes deposited  on a 
copper plate placed on block of dry ice,
as a function of time: $\rho=.4$ (crosses), $\rho=$ .8 (hexagons)}
\label{}
\end{figure}
\clearpage

The  time $t_f$  to complete the   freezing of droplets of different sizes deposited  on dry ice, and on copper and glass plates placed
on dry ice, is  shown in the tables  below.  According to the theoretical analysis, if
 the temperature $T_i$ at the base of the drop were constant,
the ratio $\rho^2/t_f$ should be a constant. Although the observed temperature varies during 
the freezing process, this constancy is   approximately valid   in the examples given in the tables below.
According to the numerical calculations of Schultz {\it et al.} \cite{schultz}, $T_i=- (.12 L/k_i )(\rho^2/t_f)$, where $L$ is the
latent heat and $k_i$ is the heat conductivity of ice. For droplets  freezing on the copper plate,  $\rho^2/t_f= .015$,  and taking  $L=333 J/cm^3$, and $k_i=.024 J/cm\; s K$,  $ T_i=-25^o$C. This value of $T_i$  is considerably higher 
than the mean value ( between $-45^o$C and $-50^o$ C ) of the time dependent temperature  during the freezing
process shown in Fig.8, while my   analytic approximation gives  $T_i= -80.5^o $ C,  which  is too low. 
For drops freezing directly on dry ice,   $\rho^2/t_f \approx .002$, and  the calculation of Schultz {\it et al.}  gives  $T_i=-3^oC$.
which  also is much higher than the mean value of the experimentally observed temperature  in Fig.7, while
my analytic approximation yields  $T_i=-10.4^o$C.  

Another problem  with the  comparisons of theory and experiment is
that  only  the smallest size droplets -  with a  radius $\rho \approx .1$  cm. - have a spherical shape, while larger droplets are 
 flattened by the effect of gravity. This effect can be  seen in the droplet shown in Figs. 1, 2.
For these droplets,  I have taken for the value of  $\rho$ one half of the diameter measured at the base of the droplet.

\begin{tabular} {| c | c | c | c | }\hline
\multicolumn{3}{|c|}{\bf I. Freezing on dry ice}\\ \hline
\multicolumn{1}{|c|}{$\rho$(cm)} &
\multicolumn{1}{|c|}{$t_f$(sec)} &
\multicolumn{1}{|c|}{$\rho^2/ t_f$} \\ \hline\hline
 .23  & 30& .0018 &  \\ \hline  
  .37   & 72 & .0019 &    \\ \hline
  .50   & 120& .0021 &  \\ \hline
  .63 &250 & .0016 &  \\ \hline
\end{tabular}

\begin{tabular} {| c | c | c | c | }\hline
\multicolumn{3}{|c|}{\bf II. Freezing on copper plate}\\ \hline
\multicolumn{1}{|c|}{$\rho$(cm)} &
\multicolumn{1}{|c|}{$t_f$(sec)} &
\multicolumn{1}{|c|}{$\rho^2/ t_f$} &\\ \hline\hline
  .37 & 9 & .015&  \\ \hline  
  .5   & 17& .015 &    \\ \hline
 .58  & 23 & .015 & \\ \hline
  .66  &  28 & .016 &\\ \hline
\end{tabular}

\begin{tabular} {| c | c | c | c | c}\hline
\multicolumn{3}{|c|}{\bf III. Freezing on glass plate}\\ \hline
\multicolumn{1}{|c|}{$\rho$(cm)} &
\multicolumn{1}{|c|}{$t_f$(sec)} &
\multicolumn{1}{|c|}{$\rho^2/ t_f$} \\ \hline\hline
.25 &29 & .0022 &  \\ \hline  
  .4   &48& .0033 &    \\ \hline
 .5   &81& .0037 &  \\ \hline
.5 & 101&.0030 &  \\ \hline
  .58 &145& .0027 & \\ \hline
\end{tabular}\\

\section*{IV. Conclusions}

I have shown that the main features of the time dependent propagation of the ice-water front of water droplets
deposited on a surface at subzero temperatures can be understood  by  Stefan's relation for planar surfaces,  together with an extension
of this relation
when this front becomes concave.  Previous   calculations of   Schultz {\it et al.} \cite{schultz}, based
on the numerical  integration of the heat diffusion differential equations for this propagation, assume that the temperature
at the base of these droplets is a constant,  while my analytic approximation can also be applied
to time varying temperatures.  My  measurements, presented in Section III,  indicate that during the freezing process this temperature  is not
constant,  and   for small drops,  $\rho \approx .1$ cm.,   it varies rapidly. 
The assumption that the theory
can   be  applied  by  taking   the  mean value of the temperature leads to some of the observed results,
e.g. the approximate constancy of the ratio $\rho^2/ t_f$ for a range of values of drop sizes and freezing times, shown
in Section III, but some of these results are  in disagreement with the direct measurements of the magnitude of this  temperature.

\section *{Acknowledgements}
I would like to thank Craig Bohren and Joshua Deutsch for helpful comments about  the application of Stefan's relation,  
 Dave Belanger for instructions on temperature measurements with a thermistor, and  M.G. Worster for his  timely suggestion 
 to freeze water  droplets on copper plates at subzero temperature.


\begin{thebibliography}{99}
\bibitem{anderson}   D.M. Anderson, M.G. Worster, and S.H. Davis, ``The case for a dynamic contact angle in containerless solidification,"
                                     Journal of Chrystal Growth {\bf 163} 329-338 (1996).
                                     
\bibitem{snow}   J.H. Snoeijer and P. Brunet, ``Pointy ice drops: How water freezes into a singular shape," Am. J. Phys. {\bf 80}, 764 (2012).

\bibitem{oscar}      O.R. Enr\'{i}quez, A.G. Mar\'{i}n, K. G. Winkels, and J.H. Snoeijer, ``Freezing singularities in water drops,"  Phys. Fluids {\bf24}
                                 09112 (2012).

\bibitem{michael}  M. Nauenberg, ``Comment on `Pointy ice drops: How  water freezes into a singular shape,'" Am.J. Phys.{\bf81}, 150 (2013).

\bibitem{snow2}   J.H. Snoeijer and P. Brunet, ``Response to Comment on Pointy ice drops: How water freezes into a singular shape," 
                                 Am. J. Phys. {\bf 81} 151 (2013).
 
 \bibitem{michael2}  M. Nauenberg, `` Conical tip in frozen water drops", arXiv: 1404.4425 v1 [physics.flu-dyn] 17 April 2014                                
                    
\bibitem{schultz}  W.W. Schultz, M.G. Worster, D.M. Anderson, `` Solidifying Sessile Water Droplets", in {\it  Interactive Dynamics of Convection and Solidification}, edited by P. Ehrhard, D.S.  Riley and P.H. Steen (Kluwer, Academic Publishers 2001) pp. 209-226.

\bibitem{stefan}   J. Stefan, `` \"{U}ber die Theorie der Eisbildung insbesonder \"{u}ber die Eisbildung im Polarmeere", S.-B Wien, Akad. Mat. Natur . 
                               {\bf 98} (1889), 965-983.  For a modern presentation of Stefan's theory,  and a large bibliography on this
                               subject, see L.I. Rubenstein, {\it The Stefan Problem}, vol. {\bf 27}
                               Translations of Mathematical Monographs, American Mathematical Society (1971)
 
 \bibitem{jacco}   A.G. Marin, O.R. Enriquez, P. Brunet, P.Colinet and J. H. Snoeijer, ``Universality of Tip Singularity Formation in Freezing Water Drops",  Physical Rev. Letters {\bf 113}, 054301 (2014)
                               
                                 
\end{thebibliography}
\end{document}